# Photoluminescence from $Cr^{3+}$ excited orbital states in magnetically ordered $CrCl_3$

Lanqing Zhou[1,2], Marjana Ležaić[3], Yuriy Mokrousov[3,4], Minh N. Bui[1,2], Renu Rani[1,5], Govind Ummethala[5], Amir H. Tavabi[5], Tatiana E. Gorelik[5], Rafal E. Dunin-Borkowski[5], Detlev Grützmacher[1,6] and Beata E. Kardynał[1,2*]

[1]Peter Grünberg Institute (PGI-9), Forschungszentrum Jülich, 52425 Jülich, Germany
[2]Department of Physics, RWTH Aachen University, 52074 Aachen, Germany
[3]Peter Grünberg Institute (PGI-1), Forschungszentrum Jülich, 52425 Jülich, Germany
[4]Institute of Physics, Johannes Gutenberg-University Mainz, 55128 Mainz, Germany
[5]Ernst Ruska-Centre for Microscopy and Spectroscopy with Electrons (ER-C), Forschungszentrum Jülich, 52425 Jülich, Germany
[6]Jülich-Aachen Research Alliance, Institute for Green-IT (PGI-10), Forschungszentrum Jülich and RWTH Aachen University, Germany

## Abstract

Control of magnetic order through optical manipulation of orbital states can be realized in a wide range of materials, including insulators. In the latter case, electrons are excited into different electronic states, whose character may be a complex result of spin-orbit-lattice coupling. Here, we use a combination of polarization-resolved photoluminescence spectroscopy and density functional theory calculations to study excited $Cr^{3+}$ ions in bulk, insulating antiferromagnetic $CrCl_3$. Our results demonstrate a temperature-dependent preferential polarization direction below ~17 K, which reorients in the presence of an applied in-plane magnetic field. Density functional theory calculations reveal a local distortion of the ligand field and the generation of out-of-plane spin and orbital moments, following $Cr^{3+}$ ion excitation. These changes link the photoluminescence polarization to the magnetic state of $CrCl_3$. Our findings provide insight into the excited-state electronic structure of $Cr^{3+}$ ions and suggest that $CrCl_3$ is a promising candidate for opto-orbitronic applications.

## Introduction

Potential applications of antiferromagnetic (AFM) materials in information storage, for example in the form of heterostructures of van der Waals materials, are resulting in interest in the optical readout of magnetic order.[1] The interplay between orbital degrees of freedom of electronic states and light is also attracting increasing attention. When a solid is subjected to a perturbation by a light pulse, the excited electronic states experience a modification of their orbital character through hybridization with other states. This process is often accompanied by currents associated with the spatial redistribution of the states and the corresponding formation of non-equilibrium magnetization states based on orbital moments, *i.e.*, orbital magnetization. Whereas aspects of orbital dynamics and currents of angular momentum are studied intensively in the steady state domain,[2, 3] the intricate processes that underlie optically-generated orbital magnetization and its interplay with spin properties are still poorly understood. It has recently been demonstrated that optical excitation of low-symmetry and magnetic materials can lead to large orbital angular momentum currents, accompanied by terahertz emission and angular momentum accumulation.[4-6] The orbital magnetization that results from such second-order processes can be sizable.[7, 8] At the same time, the generation and detection of orbital modifications of transient electronic structures in magnetic materials, accompanied by magnetization reorientation and large non-equilibrium orbital moments, remains a significant challenge.

*b.kardynal@fz-juelich.de

The longer relaxation times of electronic excitations from the d-states of transition metal ions in semiconducting $XPS_3$ (X = Mn, Ni, Fe, Co) have provided insight into the interplay between orbital, spin and lattice degrees of freedom.[9, 10] The generation of coherent phonon modes due to the displacement of ions upon excitation in $CoPS_3$ and sub-THz magnons as a result of photoinduced magnetic anisotropy in $NiPS_3$ has been studied using magnetic linear dichroism. However, the fast excitation decay did not allow insight to be obtained into the excited electronic state.[11]

Much longer excitation lifetimes have been measured in insulating magnetic chromium halides, in which electrons are strongly localized on $Cr^{3+}$ ions.[12] Photoluminescence (PL) emitted from the d-d states of $Cr^{3+}$ in $CrI_3$ and $CrBr_3$ has been shown to be circularly polarized,[13-15] with the polarization linked to the spin of electrons participating in optical transitions, although the vibronic nature of the signal did not allow the electronic and lattice degrees of freedom to be decoupled. In contrast to $CrI_3$ and $CrBr_3$, which exhibit out-of-plane magnetic order, each monolayer (ML) in $CrCl_3$ is magnetized in the plane of the Cr sub-lattice, breaking its electronic symmetry. Magnetic order in $CrCl_3$ develops from a high-temperature paramagnetic phase in two steps: 1) MLs first adopt a ferromagnetic (FM)-like phase at $T_C$= 17 K; 2) AFM coupling between successive MLs is then established at $T_N$ = 14.5 K.[16]

Here, we demonstrate that PL from a $CrCl_3$ bulk crystal, when measured at 2 K in a direction normal to the ML planes, exhibits preferential polarization direction. We study the response of the polarization to temperature and to the presence of an in-plane magnetic field. We use *ab initio* calculations to show that the observations result from the breaking of orbital symmetry in magnetized $CrCl_3$ upon electron excitation. Our results indicate that the interaction of $Cr^{3+}$ electrons with laser excitation drives the ions into an excited state with a large orbital moment, in addition to spin reorientation, thereby providing evidence for a large light-driven orbital response. Our findings provide important guidelines for applications in opto-orbitronics, where orbital generation on the sub-ns scale can result in prominent orbital transport and magnetization dynamics.

## Experimental results

Figure 1a shows representative polarization-resolved PL spectra recorded from a $CrCl_3$ bulk sample. Details of the sample and measurements are described in Methods, while the optical setup is presented in Figure S1 of the Supplementary Information (SI). The spectra were acquired at 2 K, using laser excitation at 1.8 eV. Horizontal (H) and vertical (V) polarization components were selected by using a polarization analyzer in the PL collection path. Both spectra are dominated by a broad peak between 1.27 and 1.6 eV, which is characteristic of vibronic optical transitions between crystal-field-split $Cr^{3+}$ d-orbitals (Figure 1b).[17] A weak zero phonon line (ZPL), visible at 1.575 eV, is enlarged in Figure 1a. Polarization is evident across the entire wavelength range, including the ZPL. We integrated the counts over the entire emission range (1.27-1.6 eV) to determine the degree of polarization (DoP) of the vibronic signal and used the integrated area (ZPL counts) of the fitted Gaussian over 1.567-1.584 eV to determine the DoP of the ZPL.

The total PL counts, which are dominated by the vibronic signal (1.27-1.58 eV) and ZPL counts (1.567-1.584 eV) are plotted as a function of analyzer angle in Figure 1(c-d). The data are fitted using a $\frac{(I_{max}+I_{min})}{2} + \frac{(I_{max}-I_{min})}{2} cos^2(\theta-\theta_{max})$ function, where $I_{max}$ and $I_{min}$ are the integrated counts in the relevant energy ranges, measured along $\theta_{max}$ and $\theta_{min} = \theta_{max}$ +90°, respectively. The analyzer rotation angle, $\theta$, is measured from 0°, which we assign as the H-polarization direction and is aligned with the applied magnetic field (Figure S1 and Methods section). We refer to $\theta_{max}$ (taking values from -90° to 90°) as the polarization direction and calculate the DoP using the expression $\frac{I_{max}-I_{min}}{I_{max}+I_{min}} \times 100\%$. Since the DoP of the ZPL (26.7%) is different from that of the vibronic signal (21%) in Figure 1(c-d), in the rest of the text, we focus on the ZPL to obtain insight into the electronic states that are involved in the optical transitions. We include the corresponding vibronic signal data in the SI.

According to Figure S2, the direction of polarization of PL is independent of the polarization of the excitation laser beam. This behavior suggests that there is no linear dichroism in the absorption from $^4A_2$ into the vibronic states of the $^4T_2$ electronic state and that the observed PL polarization is unrelated to the absorption.[13, 18, 19] The $^4T_2$ electronic state is the lowest-energy excited state, in which one electron is excited to the $e_g$ level, leaving a hole in the $t_{2g}$ state (Figure 1b). Without symmetry breaking in the ligand field, the hole is equally likely to occupy any of the $t_{2g}$ triplet states.[20] Even though excitons are directional,[21, 22] on average the ZPL should be unpolarized. In order to obtain insight into the mechanism of symmetry breaking that leads to polarization, we measured the samples over the temperature range between 30 K (which is above $T_C$ = 17 K for bulk $CrCl_3$) and 2 K.

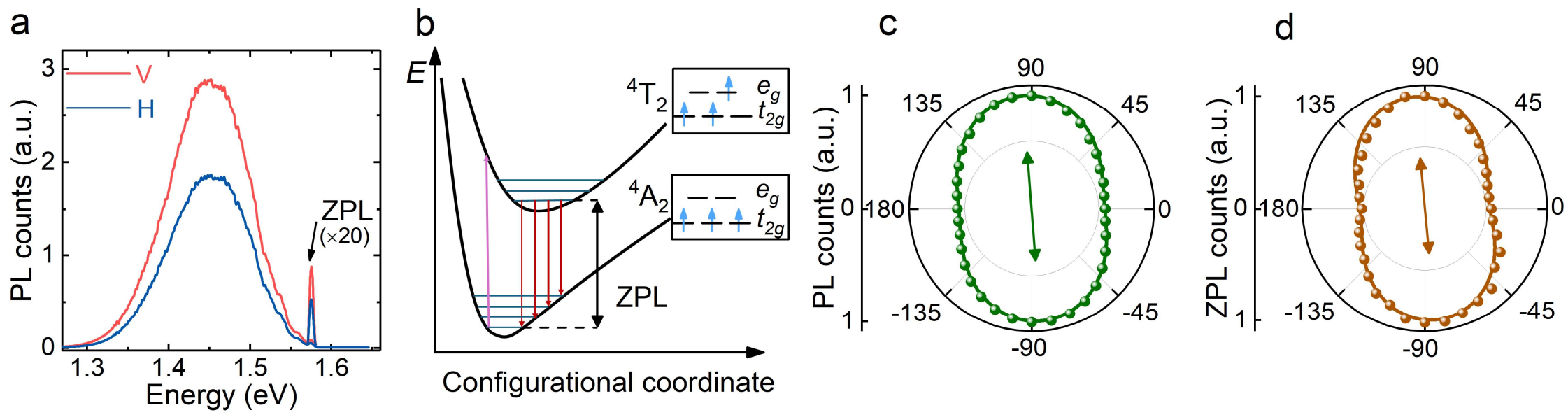

Figure 1. Photoluminescence of $CrCl_3$. (a) Vertically (V) and horizontally (H) polarized photoluminescence emission spectra from bulk $CrCl_3$ under H-polarized laser excitation. The zero-phonon line region is magnified twenty times for better clarity. (b) Schematic configurational energy diagram showing the ground ($^4A_2$) and excited ($^4T_2$) state occupations of the five 3d orbitals, which are split by the ligand field into $e_g$ and $t_{2g}$ degenerated sets. Three *d*-electrons (blue arrows) occupy three $t_{2g}$ orbitals in the ground state. One electron is excited to the $e_g$ orbital in the excited state. Following absorption (purple arrow), recombination can occur into different vibrational states of the ground state (red arrows). The zero-phonon line, which originates from a pure electronic transition, is marked with a double black arrow. (c) The integrated vibronic counts (1.27 to 1.6 eV) plotted using green symbols as a function of analyzer angle, with H along the 0° line. The counts are normalized to their fitted maximum values. The solid green line is a fitted function, with the arrow indicating the polarization axis. (d) An equivalent plot for the ZPL signal.

Figure 2 shows the representative polarization-resolved temperature dependence of the ZPL counts from one location with preferred polarization 44° at 2 K (Figure S4). The plot shows two polarization components ($ZPL_{44°}$ and $ZPL_{134°}$), along $\theta_{max}$(2 K) = 44° and $\theta_{min}$(2 K) = 134°, in the polarization basis at 2 K. The $ZPL_{44°}$ component decreases monotonically from 2 to 30 K. The $ZPL_{134°}$ component decreases slowly up to ~10 K, after which the two polarization components have similar counts. The ZPL counts measured as a function of analyzer orientation confirm that the ZPL is unpolarized at 17 K, the temperature below which the $Cr^{3+}$ magnetic moments are aligned. Similar cases at other positions are obtained as shown in Figure S5 and Figure S6, where the difference between $\theta_{max}$ and $\theta_{min}$ decrease rapidly upon warming and becomes zero above 17 K. There are variations between the temperatures at which the DoP vanishes between the different positions, but it remains below 17 K. In some cases, a residual weak polarization could be measured even at 30 K (Figure S6).

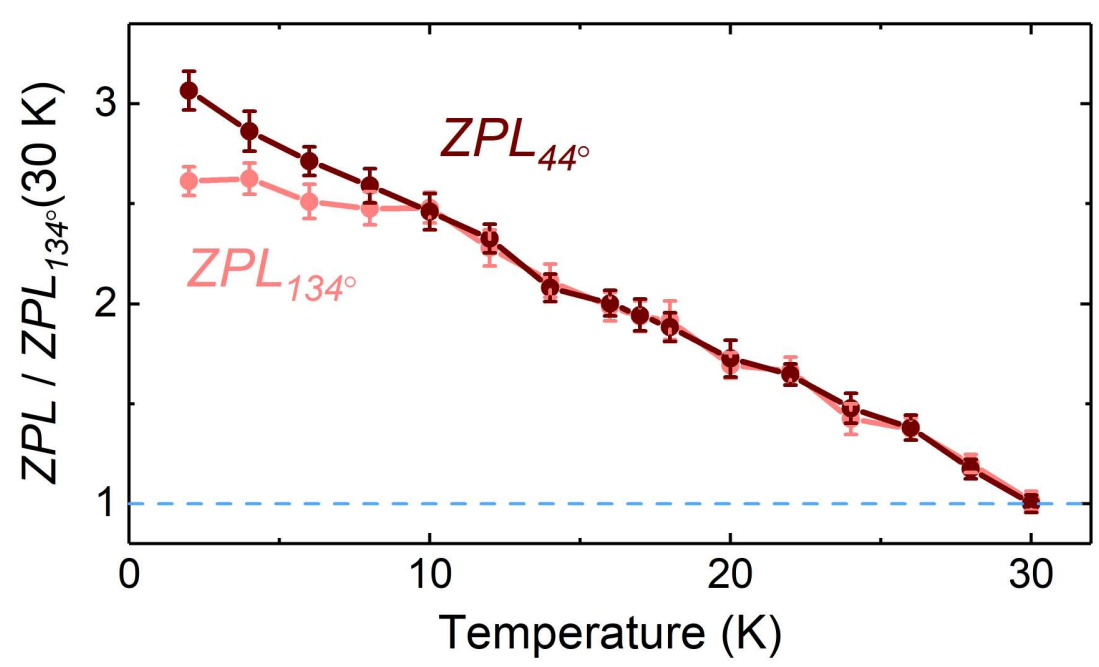

Figure 2. Temperature dependence of normalized ZPL counts for two orthogonal polarization components. The analyzer directions are fixed at $\theta_{max}$ = 44° and $\theta_{min}$= 134° measured at 2 K. The two polarization components are normalized to $ZPL_{134°}$ at 30 K for a better comparison.

The vanishing of ZPL polarization above $T_C$ in Figure 2 points to a correlation between PL polarization and magnetic order. We therefore measured PL at the same location in the presence of an in-plane magnetic field ($\mu_0$H) of up to 0.25 T. This value was chosen to be larger than the reported remagnetization field of bulk $CrCl_3$ of 0.2 T.[16, 23] Polar plots of ZPL polarization, which are shown in Figure 3(a-c), were collected at $\mu_0$H = 0, 0.05 and 0.2 T applied at 2 K along 0° (H-polarization) in the polarization reference frame. The ZPL DoP increases progressively from 0 to 0.2 T, while the polarization axis changes from 44° at 0 T to 78° at 0.05 T and then to 20° (*i.e.*, closer to the direction of the applied field) at 0.2 T. The vibronic signal undergoes similar changes (Figure S7). Figure 3d shows the evolution of ZPL counts of polarization components, measured in a fixed polarization basis defined ($\theta_{max}$ = 44° and $\theta_{min}$ = 134°) at 0 T. The 0 T maximum (minimum) component dips (peaks) between 0.02 and 0.2 T lie within the magnetic field range of the fastest magnetization changes in Figure 3e. A comparison between Figure 3d and Figure 3(a-c) shows that rotation of polarization is primarily responsible for these changes.

The measurements shown in Figure 2 and Figure 3 were repeated at several locations randomly distributed over a 160 x 240 μm area of the crystal surface and separated from each other by at least 40 μm (as shown in Figure S3). Figure 4 compares the DoP and polarization directions at 0 and 0.2 T of all the points. An increase in the DoP is evident, accompanied by a rotation towards the direction of the magnetic field (0°) in approximately half of the cases. These positions are marked in red and are referred to collectively as Group L. At 0 T, the values of $\theta_{max}$ for Group L span the light-shaded wedge from -41° to 44°. At 0.2 T, all of the polarization axes are rotated towards $\mu_0$H and the angular distribution narrows to (-22°, 20°), corresponding to the darker-shaded wedge. The other positions showed minimal rotation and are referred to as Group S. We show the polar plots of polarization at one position (Group S, $\theta_{max}$(2 K) = 17°) with the magnetic field in Figure S8. The DoPs at these positions outside (inside) the shaded wedge are visibly smaller (almost unchanged) at 0.2 T.

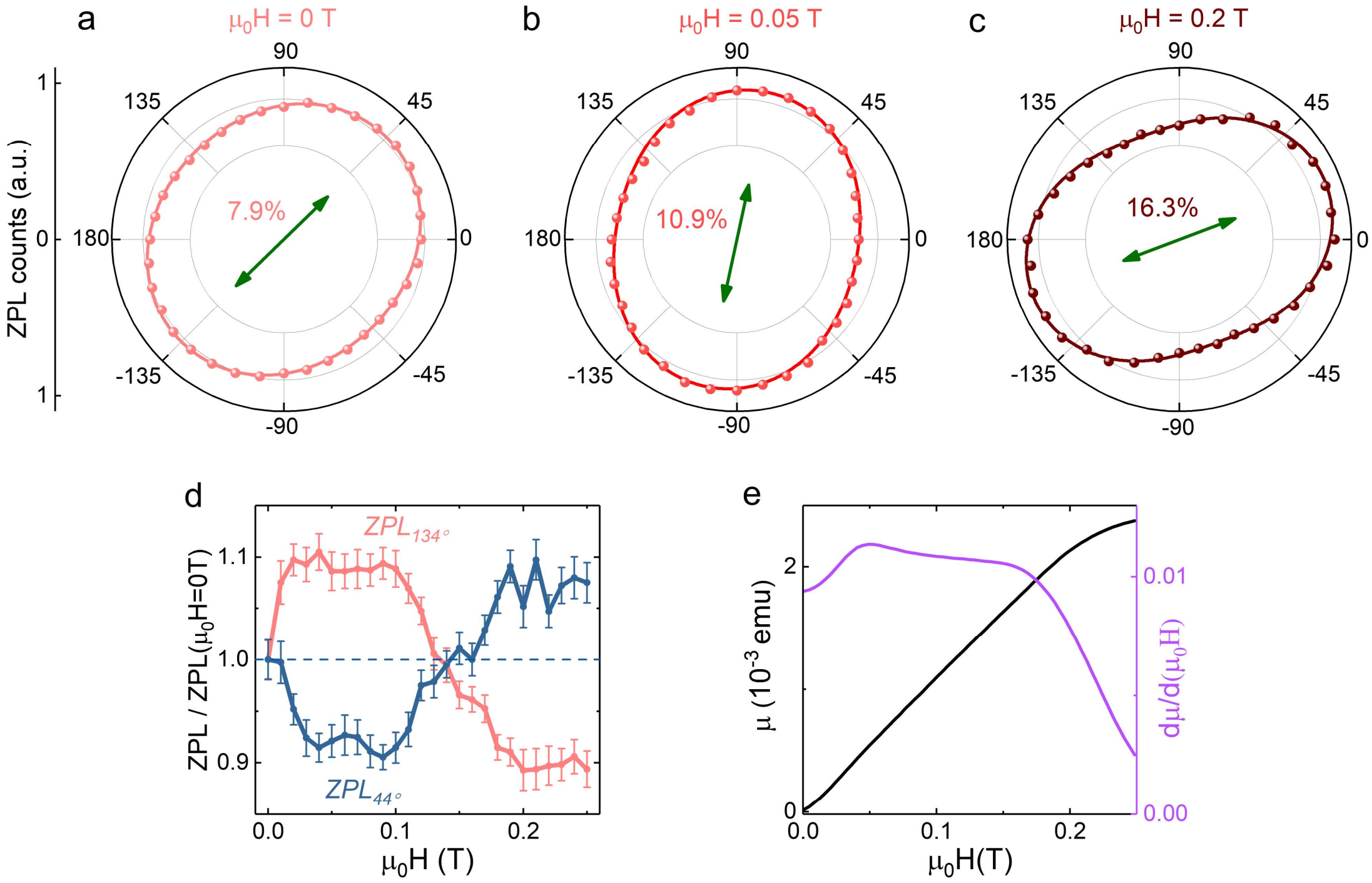

Figure 3. Polar plots of the integrated ZPL counts as a function of polarization analyzer angle at magnetic fields of 0 T (a), 0.05 T (b), and 0.2 T (c). The plots are displayed on the same radial scale, with counts normalized to the fitted maximum at 0.2 T. Solid lines show fitted functions to the data shown as points. Green arrows mark the preferred polarization directions with the percentage values of DoP given next to the arrows. The polarization angles are 44º, 78º, and 20º at 0, 0.05, and 0.2 T. (d) Evolution of two orthogonal polarization components of the ZPL at fixed polarization basis at 0 T, along 44º and 134º with the applied in-plane magnetic field ($\mu_0$H). The ZPL signals are normalized to their zero-field values. (e) The magnetization curve (black) measured at 2.5 K and its first derivative curve (purple), emphasizing the range of the fastest magnetization changes.

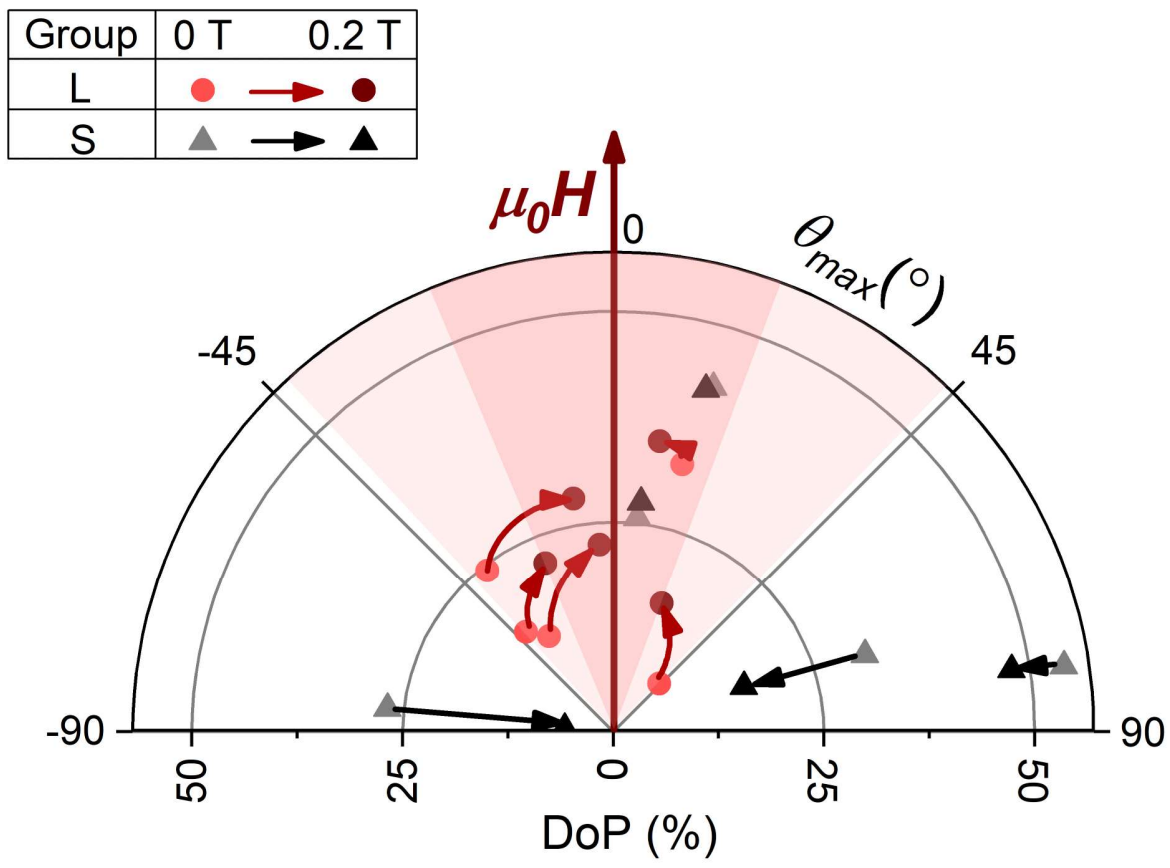

Figure 4. The legends define symbols that represent data points for each group (S and L) for the following experimental conditions: 0 and 0.2 T. Lines in the tables and in the polar representations link measurements of the same data positions under different conditions. Polar plot shows the angular distribution of polarization at 2 K for all the measured points. The direction of polarization is shown on the angular axis, with the corresponding DoP on the radial axis. The polarization for points in Group L rotates towards the applied field (along 0°). Their angular distribution (-41° to 44°), which is marked in light pink shading for 0 T, is replaced by a narrower distribution (-22° to 20°), which is marked in darker pink shading at 0.2 T. Although the DoPs for all the points in the shaded area are greater at 0.2 T, the increase from 0 to 0.2 T is minimal for the two points in Group S. The polarization of the points outside the shaded region does not rotate and the DoP decreases from 0 to 0.2 T.

## Simulations

To investigate the microscopic origin of the experimentally observed polarization of PL, we analyzed the real-space charge redistribution associated with the localized $Cr^{3+}$ excitation as well as the corresponding transition dipole matrix elements obtained from the excited-state electronic structure.

In order to simulate the excited state of $CrCl_3$, we employed the Vienna Ab-initio Simulation Package (VASP).[24-27] The $CrCl_3$ ML has a trigonal *p-31m* (L71) layer group. We performed calculations for a FM-ordered ML using a unit cell with 8 formula units. The Cr magnetic moments were aligned along the crystallographic $a$-axis (see Figure 5a), which we found to be the easy magnetization axis in the ground state. For computational details, please refer to the SI.

First, we simulate the ground state of the ion, with lattice relaxed with all the 3d electrons residing in $t_{2g}$ bands. We find that the magnetic spin moment per $Cr^{3+}$ ion is $(2.81, 0.0, 0.0)\mu_B$, which is close to the value of $3\mu_B$ expected for three electrons with aligned spins.[16, 28] The total onsite orbital magnetic moment of the $Cr^{3+}$ ion is found to be $(-0.03, 0.0, 0.0)\mu_B$, which is consistent with the expected quenching in the octahedral crystal field of the $Cl^-$ ligands.

The excited state of a $Cr^{3+}$ ion was simulated by moving one electron from the $t_{2g}$ to the $e_g$ band in the fixed-state-occupation mode of VASP followed by the lattice relaxation in accordance with Born-Oppenheimer principle. The hole in the $t_{2g}$ states and the corresponding electron in the $e_g$ states localize on one of the eight $Cr^{3+}$ cations in the supercell, resulting in a magnetic spin moment of $(2.79, 0.04, -0.88)\mu_B$ and an orbital magnetic moment of $(0.55, 0.00, -0.50)\mu_B$ for this cation.

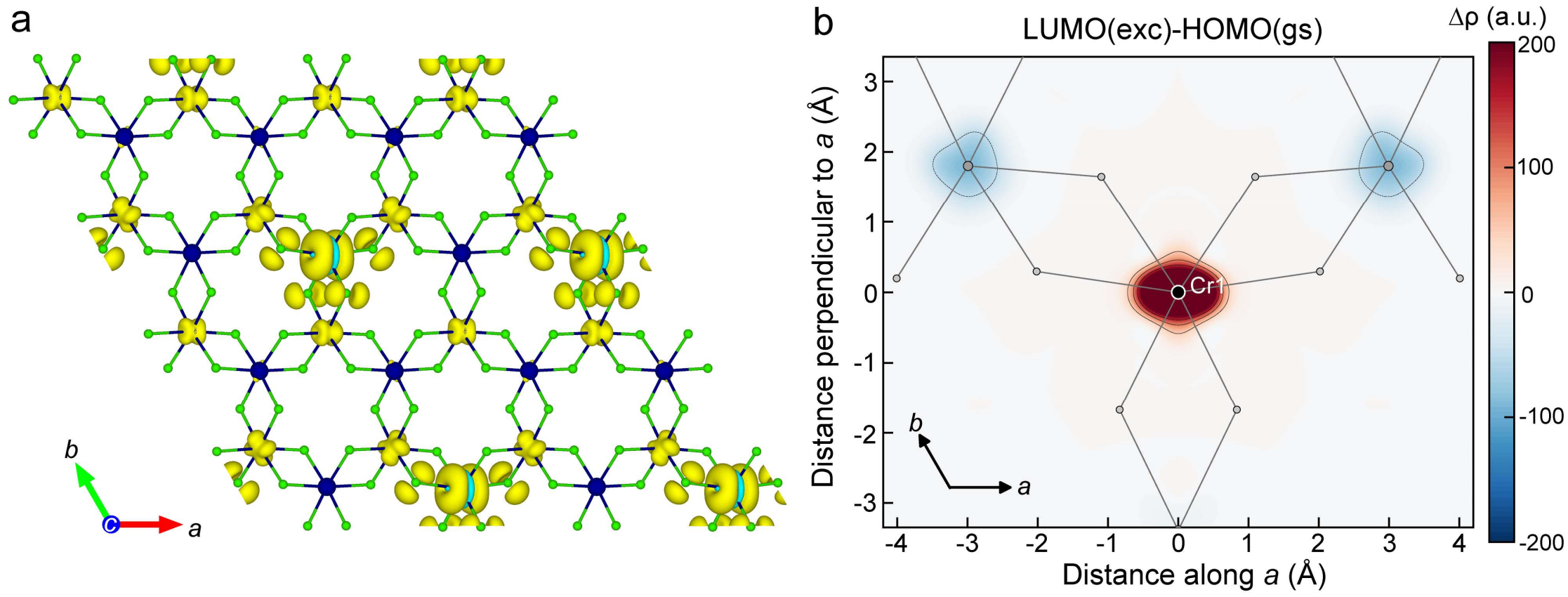

Figure 5. *Ab-initio*-calculated spatial charge distributions in a $CrCl_3$ monolayer for eight formula units and large computational unit cells, with one of the $t_{2g}$ electrons promoted to an unoccupied $e_g$ state. Dark blue (light green) spheres represent $Cr^{3+}$ ($Cl^-$) ions. (a) Spatial charge distribution of the hole in the $t_{2g}$ band of a magnetically ordered $CrCl_3$ monolayer, with the spins pointing along the crystallographic *a*-axis. (b) Real-space charge density difference of the LUMO of the excited state (exc) and HOMO of the ground state (gs), both at the relaxed excited state atomic positions, in the *ab*-plane around the excited Cr1 site. Gray circles correspond to the projections of the positions of the surrounding atoms onto the *ab*-plane. The pronounced anisotropy along the crystallographic *a*-axis is consistent with the observed linear polarization of the PL emission. The color scale is shown in arbitrary units.

The large orbital moment, which points along $Cr^{3+}$-$Cl^-$ bonds with an in-plane projection along the magnetization direction (having components along the *a*-axis as well as out-of-plane), is a consequence of the hole occupying the state, which is a linear combination of the $t_{2g}$ states of the $Cr^{3+}$ ion, with the characteristic charge distribution shown in Figure 5a. The symmetry of the ML near the excited cation is reduced to a monoclinic *c211* (L10) layer group. This distortion mostly affects the positions of the Cl atoms, leaving the Cr sublattice almost unchanged. In this way, the lattice parameters of the monolayer 2D crystal essentially stay the same, but the symmetry of the ligand field around the excited $Cr^{3+}$ ion is reduced.

The excitation breaks the initial symmetry of the hexagonal lattice existing in the ground state. Since atomic bonding can change upon electronic excitation, we performed atomic relaxation in this state, resulting in a contraction in bond length between the excited $Cr^{3+}$ ion and its first $Cr^{3+}$ neighbor (in its ground state) in a direction perpendicular to the magnetization direction, resulting in a monoclinic unit cell for the excited state. (The corresponding cif file is provided in the SI).

To gain insight into the radiative recombination channels, we simulate the ground state in the atomic configuration of the excited ion as the lifetime of the electron in the excited state is much longer than the lattice relaxation time.[12] Following deexcitation, the $Cr^{3+}$ ion is occupied with three electrons in the $t_{2g}$ states, the orbital moment vanishes but the state is slightly rotated compared with the relaxed ground state.

Figure 5b shows the charge-density redistribution associated with the excitation, constructed from the difference between the lowest unoccupied molecular orbital (LUMO) of the excited state and the highest occupied molecular orbital (HOMO) of the ground state, both in the lattice relaxed in the excited state. The redistribution exhibits a pronounced in-plane anisotropy, with the dominant weight elongated along the crystallographic *a*-axis. While the charge-density difference itself does not uniquely determine the polarization direction, it already suggests a substantial lowering of the local in-plane symmetry around the excited Cr ion and indicates that the optical transition is not isotropic within the layer plane.

To quantify the polarization properties, we evaluated the transition dipole matrix elements between the occupied and unoccupied excited-state orbitals at the dominant k-points contributing to the localized excitation. The analysis was performed for the k-points exhibiting the strongest simultaneous localization of both the hole and electron components on the excited Cr ion. For all dominant k-points, the transition dipole matrix element was found to be overwhelmingly polarized along the Cartesian x-direction, which in the present lattice convention coincides with the crystallographic *a*-axis (i.e. with the direction of the applied magnetic field).

Specifically, the calculated ratios of the in-plane dipole intensities $|d_x|^2/|d_y|^2$ amount to approximately $3.9\times10^2$, $3.8\times10^2$, $2.4\times10^2$, and $3.3\times10^2$ for the four leading k-points. Thus, the optical transition intensity along the *a*-axis exceeds the perpendicular in-plane component by roughly two to three orders of magnitude throughout the relevant region of reciprocal space. The out-of-plane component remains comparatively weak, with $|d_z|^2/|d_x|^2 \sim 0.06–0.08$.

Importantly, although the complex phases of the dipole matrix elements vary among different k-points, the corresponding dipole intensities remain highly anisotropic and robust, indicating that polarization anisotropy is an intrinsic property of the excited-state manifold rather than an accidental feature of a single k-point.

## Discussions

The simulation results provide a direct microscopic explanation for the observed PL polarization. They demonstrate that the localized excited state possesses a strongly anisotropic transition dipole moment whose dominant component is aligned with the direction of spin moment of the ion. It is important to emphasize that magnetization direction only defines the direction of anisotropy in the excited state, the matrix element is not directly dependent on the spin.

The observed degrees of polarization are not as high as the anisotropies of the calculated matrix elements. However, we emphasize that the present computational approach does not constitute a full many-body optical calculation. In particular, the employed transition dipole matrix elements are obtained from reconstructed pseudo-wavefunctions and therefore do not include the full projector-augmented-wave (PAW) optical matrix elements, excitonic effects, or explicit electron-hole interactions. Consequently, the present analysis should be interpreted as a microscopic and symmetry-resolved characterization of the excited-state transition anisotropy rather than as a quantitatively rigorous prediction of absolute optical intensities. A formally complete treatment of the PL transition matrix elements would require, for example, time-dependent density-functional theory (TDDFT), Bethe-Salpeter-equation (BSE), or related many-body approaches including the full optical transition operator.

The simulations predict the emission from an individual ion. Since emission is collected from all ions in the volume of the focused laser beam, our conclusions about directional polarization of PL apply to a crystal that is magnetized uniformly along the crystallographic *a*-axis in that volume. In such a case, the spin moments of $Cr^{3+}$ ions are aligned, the monoclinic distortion around randomly excited $Cr^{3+}$ cations has the same orientation throughout a magnetic domain, and the PL from the domain is polarized. Above the Néel temperature, magnetic moments of $Cr^{3+}$ ions are random and polarization is not expected in the simulation and is not observed. Similarly, the observed rotation of polarization in the external magnetic field is explained by the simulations.

The magnetization direction of the crystal can be found by minimizing the micromagnetic energy, which for $CrCl_3$ in an external in-plane magnetic field comprises two terms: the energy of interactions between the magnetization vectors of neighboring monolayers and the energy of interactions between the magnetization vectors of the monolayers and the magnetic field.[29] While a unique solution can be found for 2-3 monolayer-thick films (see Note III and Figure S9 in the SI), this is not the case for thicker layers, especially in the presence of lattice defects which can locally pin the magnetization. However, according to Wang *et al.* [29], in the limit of bulk $CrCl_3$, spin flop should occur as soon as the field is applied ($H$=0), while spin flip should take place at $H$=2$H_{ex}$, in good agreement with our and other magnetization measurements.[29-31] It is also noticeable that the changes to polarization in Figure 3d occur in the same range of magnetic fields as the fastest remagnetization of the $CrCl_3$ crystal shown in Figure 3e. Considering that the volume of $CrCl_3$ probed in magnetization measurements is orders of magnitude larger than in PL and the PL exhibits non-homogeneous behavior (Figure 4), the two results are consistent with each other.

At 0.2 T, the polarization axes in the shaded wedge in Figure 4 are distributed between -22° and 20°. If we assume a uniform distribution of $\theta_{max}$ in the wedge spanning between ±22°and that $\theta_{max}$ reflects the magnetization direction, then the average magnetization per ion is $0.987\mu_i$, where $\mu_i$ is the magnetic moment of one ion in the ground state (and most ions are unexcited). This value is much larger than the magnetization at 0.2 T, which is 0.77 of that at 2 T (Figure S10). The average magnetization per ion for all points in Figure 4 at 0.2 T is $0.732\mu_i$, which is consistent with the magnetization measurements.

The origin of the behavior of measurements in group S is unclear. However, we hypothesize that it may be due to the lattice remaining in the high-temperature monoclinic phase ($\mathrm{C2/m}$) in parts of the crystal (stacking faults) at low temperature.[32-35] The high-temperature phase has ligand field symmetry of $C_{2h}$, irrespective of the magnetic order.[36] The $Cr^{3+}$ ion lacks $C_3$ symmetry and group theory predicts directional polarization of optical transitions in this case. This structural symmetry breaking could therefore explain the polarization above 17 K in group S (Figure S6). The critical field for $CrCl_3$ in the monoclinic phase has been shown to be tenfold larger than that in the low-temperature rhombohedral phase ($\mathrm{R\overline{3}}$).[29, 30, 37] Notice that we observed a similar enhanced field (Figure S10) in magnetization measurement, which may result from the presence of the high-temperature monoclinic phase. Electron diffraction from thick exfoliated samples taken from the same batch of crystals confirmed the presence of stacking faults at room temperature (Figure S11), in agreement with other reports.[32-35] This means that we measured the points in group S only during the early stages of remagnetization, when rotations of the spin moment cannot be easily predicted, especially in the low-symmetry crystal phase. The strong reduction in polarization in the three cases where the initial polarization directions are close to the normal to the applied magnetic field (Figure 4) could also arise from competing symmetry breaking between the two mechanisms. Given that stacking faults are random,[34, 38, 39] they could also explain the variations in the behaviors between the measured points.

Uniaxial strain could, in principle, also explain our observations, if it distorted the ligand field around $Cr^{3+}$ ions in competition with magnetization. Strain has been reported to cause stacking faults,[32, 40, 41] and could lead to the presence of yet different stacking orders (low symmetry) besides the low-temperature rhombohedral phase.[42]

PL in some magnetic insulators has been shown to originate from impurities as a result of efficient diffusion of free excitons to the impurity sites.[43, 44] An incoherent diffusion of excitons that are self-trapped by lattice deformation at 1.6 K as in our case is, however, expected to be low since phonons and magnons necessary for the process have very low populations, while the distortion of the lattice and large Stokes shift (250 meV) render coherent tunneling improbable.[45] Excitons could diffuse before the electronic and lattice relaxation, soon after laser excitation. Both the electronic and lattice relaxation are expected to be of the order of a picosecond or shorter,[46, 47] so we expect the diffusion to be negligible at such short times.[48] Short diffusion length in $CrI_3$ has recently been reported.[48] A fraction of excitons decaying from the impurities could explain the variable DoP observed in our experiments as impurities emission is likely to be unpolarized (defects are introduced at random sites[49-51] or in Cr-sublattice voids[52] and should not affect the "global" $C_3$ symmetry.

We would also like to comment on the consequences of spin- and orbital-moment differences between the ground and excited states. Our simulations have indicated that a hole in the excited state acquires an out-of-plane spin and orbital moment component. Refilling this state with an electron during radiative recombination creates a spin and orbital-moment defect in the sea of aligned spins of the $Cr^{3+}$ sub-lattice. Relaxation of that spin could generate a magnon, whose properties would be linked to the direction of magnetization and entangled with the generated photon. Similarly, a loss of orbital angular momentum may take place via the generation of orbital currents or conversion into angular momentum of magnons and phonons, possibly mediating magnon-phonon interactions and resulting in the formation of magnon polarons. It also suggests that the ZPL does not originate from a purely electronic transition, but rather from one accompanied by the generation or annihilation of a magnon. Further direct studies of orbital- and spin-moment changes are necessary to confirm the findings.

## Conclusions

We have shown that photoluminescence from a $CrCl_3$ bulk crystal exhibits a preferential polarization direction below the magnetic phase transition temperature. The observed polarization behavior varies across the crystal with an applied in-plane magnetic field. It either reorients or diminishes without rotation in an applied in-plane magnetic field up to 0.2 T.

Density functional theory calculations reveal that a hole generated in the $t_{2g}$ state of $Cr^{3+}$ by electron excitation into the $e_g$ state leads to a local monoclinic distortion of the ligand field along the spin direction of the $t_{2g}$ electrons of the excited cation. This distortion is shown to result in strongly anisotropic matrix elements for radiative recombination along the $t_{2g}$ spin direction. These results explain the appearance of polarization below 17 K and its reorientation in an applied in-plane magnetic field. Further, the measured distribution of the polarization directions across the sample at the applied magnetic field of 0.2 T qualitatively agrees with the magnetization measurements of this crystal. In contrast, stacking faults (e.g., the presence of high-temperature phase of $CrCl_3$) and temperature-dependent strain in the monolayer could break the symmetry of the ligand field and explain the measurements in which polarization persists above the critical temperature and is not reoriented with the applied magnetic field. This hypothesis should be tested in the future. Comparing the findings of simulations of light absorption in $CrCl_3$ with our results suggests that simulations of PL should begin from an atomic configuration of the excited state and realistic magnetization.

Remarkably, our simulations indicate that the magnetic orbital and spin moments of the hole in the $t_{2g}$ state of the excited $Cr^{3+}$ ion are not collinear with the magnetization but have an out-of-plane component. Subsequent relaxation of electronic excitation may create spin- and orbital-moment defects, which should be explored further, as $CrCl_3$ could become a model system for the heralded generation of magnons and magnon-polarons. This point may be relevant for quantum information, as magnons are considered as preferred candidates for coupling between optically driven spin qubits, and as such should be studied further.

## Methods

### Experimental details

The investigated polycrystal, as-purchased from 2D Semiconductors, was cut into a $3 \times 3$ mm$^2$ flake of thickness exceeding 100 μm and fixed to an $SiO_2$/Si substrate using cryo-stable glue. The sample, affixed to a copper plate, was secured to an *x-y-z* piezo positioner, enabling reliable nano-positioning over a 5 mm range with approximately 10 nm resolution at 2 K.

The assembly was placed in an Attodry2100 cryostat sample space filled with He gas on an insert with an optical window, facilitating light coupling for *in-situ* optical characterization. It was cooled to 10 K in 2.5 hours. It took another 10 hours to stabilize the temperature at its lowest point. Once cold, the temperature of the crystal varied between 2 and 30 K during consecutive measurements.

Measurements at 2 K were conducted at several random positions on the sample spanning a 160 μm × 240 μm rectangle. Their distribution is shown in Figure S3. Each spot is marked with its polarization angle in zero magnetic field, along with its assigned group (L or S). At each position, the PL signal was maximized by adjusting the position of the objective lens above the crystal using a *z*-positioner. At each position, spectra were collected as a function of polarization detection angle. The positions were located at least 1.3 mm from the edges, separated from each other by at least 40 μm and feature a well-defined zero-phonon signal. Their relative positions were obtained by recording their *x* and *y* coordinates, as measured by the positioners' readout.

Polarization-resolved magneto photoluminescence measurements were performed in Voigt geometry with a magnetic field applied parallel to the $CrCl_3$ crystal monolayers (in-plane), and PL excited and collected normal to the layers. The 0º (H-polarization direction) analyzer rotation angle is defined along the applied magnetic field.

The optical head, which is shown schematically in Figure S1, utilizes fiber coupling to deliver laser light and collect polarized light to a Czerny-Turner spectrometer. A low excitation power of 400 nW was maintained to avoid heating. A 40-fold increase in PL counts was observed when the laser power was increased, indicating that the excitation level of $Cr^{3+}$ was low in our experiments. Polarization control and analysis were performed in a free-space optical path to avoid depolarization in fibers. The continuous wave laser beam, with a photon energy of 1.8 eV, was linearly polarized in the back focal plane of the objective lens, which was an aspheric lens with a numerical aperture of 0.68. Linear polarization was set by a half-wave plate (HWP) after a linear polarizer (LP). Any ellipticity was corrected with a quarter-wave plate (QWP).

The same objective lens was used to collect PL. Polarization measurements were facilitated by adding a polarization analyzer composed of a half-wave plate ($HWP_2$), followed by a linear polarizer in front of the optical fiber. The rotatable $HWP_2$ serves as an analyzer of the polarization of the emitted light, while $LP_2$ ensures that all the light has the same polarization orientation before entering the fiber, thereby avoiding the polarization sensitivity of the optical components beyond this point. A 700 nm low-pass filter was placed on the input of the spectrometer to remove any laser light excitation.

**Ligand field $^4T_2$ photoluminescence of chromium trihalides.**

Chromium trihalides ($CrX_3$, X = I, Br, Cl) form a family of layered, insulating compounds. Each monolayer of $CrX_3$ is ordered ferromagnetically at low temperature, while the monolayers are coupled ferromagnetically or antiferromagnetically depending on the halide atoms.[53, 54] The octagonal crystal field of the halide ligands splits the d-orbitals of $Cr^{3+}$ into a lower energy triplet and a higher energy doublet. $^4A_2$, with three spin-aligned electrons occupying a low-energy $t_{2g}$ triplet of d-orbitals, is the ground state of the $Cr^{3+}$ ion in $CrX_3$. The lowest-energy excited state is $^4T_2$, in which one electron is lifted to the $e_g$ doublet. (See the energy diagram in Figure 1b).[13, 17, 46, 47]

**Calculation details**

Density functional theory (DFT) was employed for electronic structure calculations, as implemented in the Vienna Ab-initio Simulation Package (VASP)[55-58] with the Projector Augmented-Wave Method.[55] The cut-off energy for the plane-wave basis was set at 500 eV. A Perdew-Burke-Ernzerhof generalized gradient approximation revised for solids (PBEsol)[56] was used for exchange and correlation. Enhanced correlation effects in the Cr d-shell were accounted for within the DFT+U method using the Liechtenstein approach,[57] with values of the screened on-site Coulomb repulsion U = 1.79 eV and the intra-atomic exchange interaction J = 0.85, as evaluated in Ref.[58], within the constrained random phase approximation.[59] The Brillouin zone was sampled on a 5 × 5 × 1 k-point mesh. Spin-orbit coupling effects were included. We simulated a ferromagnetically-ordered monolayer of $CrCl_3$ in a repeated slab geometry using a calculation unit cell containing eight formula units per monolayer and a ~20 Å thick vacuum region that prevents spurious interaction effects between the simulated monolayers.

After atomic relaxation in the excited state, we obtained a monoclinic supercell, which is provided in SI in .cif format. Note that relaxation was performed for a hexagonal unit cell containing eight Cr atoms (*i.e.*, a 2 × 2 supercell) starting from optimized ground-state positions. The symmetry of the final unit cell of the excited state was determined (and the corresponding cif file created) by the FINDSYM utility.[60, 61]

**Electron diffraction**

3D electron diffraction (3D ED) data from exfoliated $CrCl_3$ crystals revealed a centered monoclinic unit cell with lattice parameters a = 6.0098 Å, b = 10.3460 Å, c = 6.1445 Å, $\alpha = 90°$, $\beta = 109.747°$, $\gamma = 90°$, in good agreement with the bulk structure, indicating that the individual layers retained their structural integrity. When the crystals were oriented with the [001] direction along the incident electron beam direction, they exhibited a typical diffraction pattern with a hexagonal metric and a 2mm intensity distribution (Figure S10a). Lines of diffuse intensity (Figure S10b) running along the c* direction, which are characteristic of the presence of stacking faults, were observed in [100] zone-axis diffraction patterns.

The diffuse scattering showed an extinction pattern, with each third row of reflections appearing discrete. This effect arises from the layer shift vector in the faulted stacking sequence, which has a magnitude of one-third of the lattice periodicity. The spacing between the reflections along the discrete rows matched the c lattice parameter of the bulk structure, indicating that the interlayer distance remains unchanged at 5.8 Å.

Since the crystals studied contain stacking faults, any analysis of the symmetry becomes obsolete. As noted above, the presence of a layer distortion due to excitation of an electron on a $Cr^{3+}$ ion does not significantly change the lattice parameters of the ML but primarily alters its symmetry. Whether such a distortion of an ML would be visible experimentally using electron diffraction remains an open question. A diffraction pattern recorded at normal incidence (orthogonal to the layer surface) is dominated by the Cr sublattice, which remains intact primarily upon distortion. However, the positions of the Cl atoms modify the intensities of weaker reflections (particularly a group of reflections around 2.97 Å), including Bragg spots that differ due to the reduced layer symmetry. The relative intensities of these reflections could serve as an indication of symmetry reduction.

**Magnetization measurement**

The magnetic moment of the $CrCl_3$ bulk crystal was measured using vibrating sample magnetometry (VSM) integrated into a Quantum Design Dynacool Physical Property Measurement System (PPMS). The sample was mounted on a rod that vibrates at a specific frequency in an in-plane magnetic field (H‖ab). The vibration induces an oscillating magnetic flux in nearby pickup coils, thereby generating an electrical signal that is proportional to the magnetic moment of the sample. In this technique, the sample magnetization is detected as an electrical current induced in pickup coils when the sample is moving up and down along the axis of the coil. Figure 3e and Figure S9 present measurements of magnetic moment *vs* magnetic field, which are consistent with the spin alignment process.[19, 25, 34]

## Acknowledgements

L.Z. acknowledges funding from the China Scholarship Council project no. 20173109 "Photon sources from two-dimensional materials". M.L. acknowledges computing time on the supercomputer JURECA[32] in Forschungszentrum Jülich under grants JIFF38, JIFF40 and REPSMANOM. Y.M. acknowledges financial support from the Deutsche Forschungsgemeinschaft (DFG, German Research Foundation) — TRR 288/2 -- 422213477 (project B06), TRR 173/3 – 268565370 (project A11). This work was supported by the EIC Pathfinder OPEN grant 101129641 "OBELIX". M.N.B. and B.K. acknowledge funding from the Volkswagen Foundation, project no. 93425, within the "Integration of Molecular Components in Functional Macroscopic Systems" initiative. R.R. acknowledges funding from the Humboldt Association. We acknowledge the Helmholtz Nano Facility staff in Forschungszentrum Jülich for their assistance in fabricating the substrates.

## Author contributions

L.Z. and B.K. planned the project, performed the experiments and wrote the manuscript with significant inputs from M.L., Y.M. and R.E.D.-B. M. L. performed the calculations. M.N.B. and D.G. supported optical data collection. R.R. conducted magnetization measurements. G.U., A.H.T., T.E.G. and R.E.D.-B conducted electron diffraction measurements. All authors reviewed the manuscript.

## Competing interests

The authors declare no competing interests.